\documentclass[conference]{IEEEtran}

\usepackage{xspace,amsmath,amssymb,amsfonts,epsfig,subfigure,syntonly}
\usepackage{cite,bm,color,url,textcomp,empheq,boxedminipage}
\usepackage{algorithmicx,algorithm}
\usepackage{epstopdf,makecell}
\usepackage{empheq}
\usepackage{pifont}

\usepackage{graphicx,graphics}  
\usepackage{multirow,multicol}
\usepackage{psfrag}    
\usepackage{stfloats}
\usepackage{url}

\newcommand{\mv}[1]{\mbox{\boldmath{$ #1 $}}}

\long\def\symbolfootnote[#1]#2{\begingroup
\def\thefootnote{\fnsymbol{footnote}}
\footnote[#1]{#2}\endgroup}

\psfull

\allowdisplaybreaks[1]

\begin{document}

\title{Mobile Edge Computing for Cellular-Connected UAV: Computation Offloading and Trajectory Optimization}
\IEEEspecialpapernotice{(Invited Paper)}
\author{\IEEEauthorblockN{Xiaowen~Cao\IEEEauthorrefmark{1},
Jie~Xu\IEEEauthorrefmark{1},
and
Rui~Zhang\IEEEauthorrefmark{2}}
\IEEEauthorblockA{\IEEEauthorrefmark{1}School of Information Engineering, Guangdong University of Technology, Guangzhou, China}
\IEEEauthorblockA{\IEEEauthorrefmark{2}Department of Electrical and Computer Engineering, National University of Singapore, Singapore}
Email: caoxwen@outlook.com,~jiexu@gdut.edu.cn, elezhang@nus.edu.sg
}

\markboth{}{}
\maketitle

\setlength\abovedisplayskip{0.1pt}
\setlength\belowdisplayskip{0.1pt}




\begin{abstract}
This paper studies a new mobile edge computing (MEC) setup where an unmanned aerial vehicle (UAV) is served by cellular ground base stations (GBSs) for computation offloading. The UAV flies between a give pair of initial and final locations, during which it needs to accomplish certain computation tasks by offloading them to some selected GBSs along its trajectory for parallel execution. Under this setup, we aim to minimize the UAV's mission completion time by optimizing its trajectory jointly with the computation offloading scheduling, subject to the maximum speed constraint of the UAV, and the computation capacity constraints at GBSs. The joint UAV trajectory and computation offloading optimization problem is, however, non-convex and thus difficult to be solved optimally. To tackle this problem, we propose an efficient algorithm to obtain a high-quality suboptimal solution. Numerical results show that the proposed design significantly reduces the UAV's mission completion time, as compared to benchmark schemes.
\end{abstract}

\vspace{-0cm}
\section{Introduction}\label{sec:intro}\vspace{-0cm}
With recent technology advancement and manufacturing cost reduction, unmanned aerial vehicles (UAVs) have received growing interests in various applications such as cargo delivery, filming, rescue and search, etc \cite{yongzeng16UAVsurvey}. 
To maintain the UAVs' safe operation with real-time command/control and enable their new applications with artificial intelligence (AI), it becomes increasingly important to enhance the communication and computation capabilities of UAVs. In order to provide reliable communication for UAVs, cellular-connected UAV communication has recently emerged as a viable new solution, in which UAVs are integrated into cellular networks as new aerial mobile users \cite{shuowenzhang17,Huangyw17}. As compared to the conventional direct UAV-to-ground communication with limited range \cite{yongzeng16UAVsurvey}, the cellular-connected UAV communication is able to provide seamless wireless communication for UAVs. By contrast, there has been very limited work addressing how to improve the computation performance of UAVs. Notice that in the forthcoming AI era, UAVs need to handle computation-intensive and yet latency-critical tasks, while in practice they usually have limited computation resources on-broad due to their size, weight, and power (SWAP) limitations. Therefore, it is imminent as well as challenging to solve the open problem of how to significantly enhance the computation power for future UAVs.

To tackle the above challenge, this paper proposes a new approach by jointly exploiting the techniques of mobile edge computing (MEC) and cellular-connected UAV communication. With MEC, cloud-like computing functionalities are provided at the edge of wireless networks such as cellular base stations (BSs) \cite{Mao17}. As a result, UAVs with cellular connection can offload their intensive computation tasks to ground BSs (GBSs) for remote execution. As GBSs are nowadays deployed almost everywhere, this new approach can provide both seamless communication and ubiquitous computation services for UAVs, which help increase their operation range and enlarge their application horizon.

The new setup of MEC with high-mobility UAV users poses new opportunities as well as challenges for the optimal computation offloading design. First, as compared to the traditional mobile user with complex fading channel with its associated GBS, a UAV user in the sky usually possesses stronger and more reliable line-of-sight (LoS) links with a large number of GBSs at the same time. This thus enables each UAV to simultaneously connect with multiple GBSs to exploit their distributed computing resources to improve the computation capability. Second, since the UAV has controllable mobility in the three-dimensional (3D) airspace, its trajectory can be jointly designed with its scheduling of computation offloading to the GBSs associated along the trajectory to optimize the performance. This is considerably different from prior studies on MEC with communication and computation resource allocation at a fixed terrestrial user and its associated GBS only (see, e.g., \cite{Bar14,Cao17,Wang17,MHchen16CAP}), thus deserving a dedicated new investigation.

Specifically, this paper considers a practical scenario where a UAV is designated to fly from an initial location to a final location, during which it needs to accomplish certain computation tasks. We assume that the UAV can arbitrarily partition these tasks into smaller-size subtasks, and offload them to some selected GBSs along its trajectory for parallel execution. Under this setup, we aim to minimize the UAV's mission completion time or total flight duration by jointly optimizing its trajectory and computation offloading scheduling, subject to the maximum speed and initial/final location constraints of the UAV, as well as the GBSs' individual computation capacity constraints. Although the formulated problem is non-convex and difficult to be solved optimally, we propose an efficient algorithm to obtain a high-quality suboptimal solution by using the techniques of alternating optimization and successive convex approximation (SCA). Numerical results show that the proposed design significantly reduces the mission completion time for the UAV as compared to other benchmark schemes.


It is worth noting that there has been  prior work \cite{sj17} that investigated another type of UAV-MEC system, where the UAV is employed as a moving MEC server in the sky to help execute the computation tasks offloaded by multiple ground users. By contrast, this paper studies a new and different scenario where the UAV is the mobile user that offloads computation tasks to multiple GBSs.
\vspace{-0cm}
\section{System Model}\label{sec:system1}
\vspace{-0cm}
\begin{figure}
\centering
 \setlength{\abovecaptionskip}{-4mm}
\setlength{\belowcaptionskip}{-4mm}
    \includegraphics[width=3.5in]{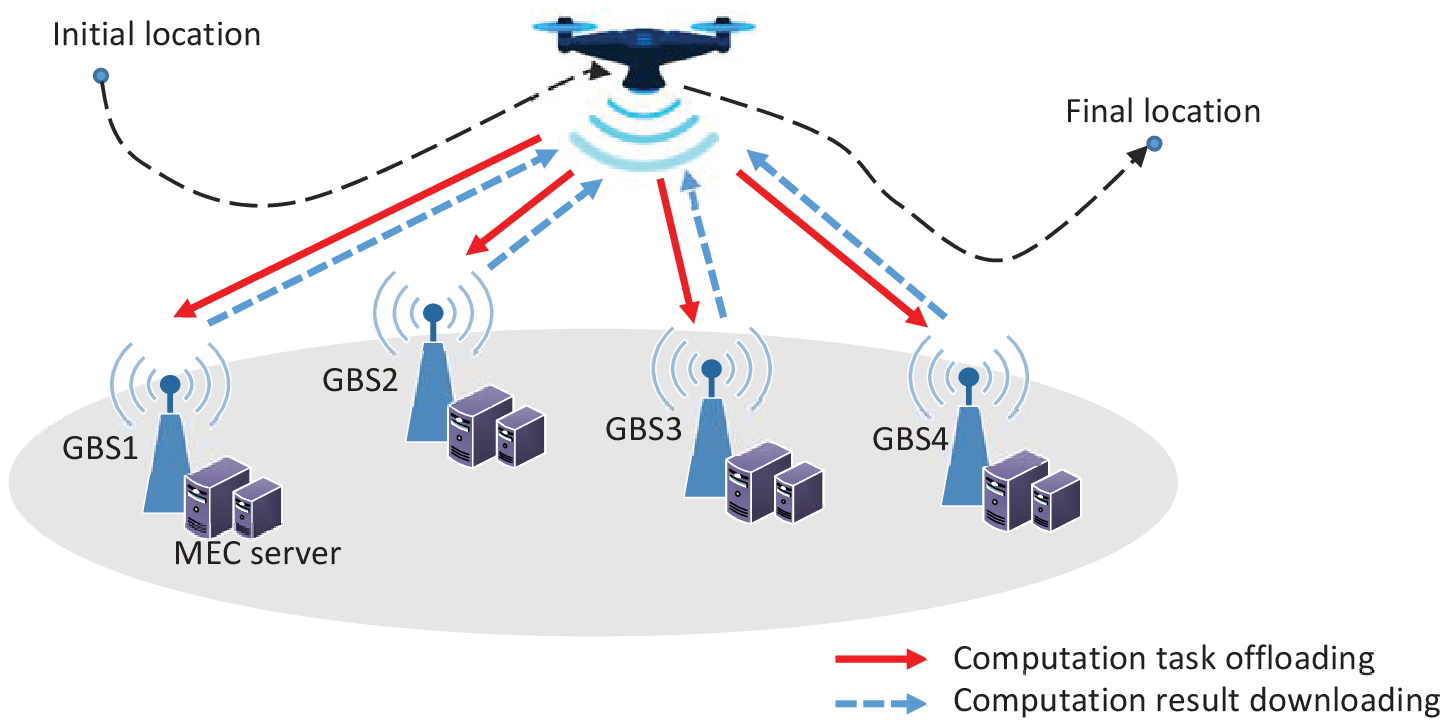}
\caption{Illustration of the new MEC scenario where a cellular-connected UAV is served by multiple GBSs along its trajectory for computation offloading.} \label{fig:1}
\vspace{-0.8cm}
\end{figure}

In this paper, we consider a new MEC system with one single cellular-connected UAV user and a set $\mathcal {K} \triangleq \{1,\ldots,K\} $ of $K\ge 1$ GBSs with MEC functionality. The UAV has a mission to fly from an initial location to a final location, during which it also needs to accomplish certain computation tasks by offloading them to the GBSs for remote execution.{\footnote{Due to the SWAP limitations, the UAV usually has limited local computation resources. In this case, we consider that the UAV user does not perform any local computing, for the purpose of exposition. }}
Let $L$ denote the number of task-input bits. We assume that the computation tasks can be arbitrarily partitioned into smaller-size subtasks that can be offloaded to different GBSs and executed in parallel \cite{Mao17}.
We further assume that at each GBS $k \in\mathcal K$, the execution of each task-input bit requires the same number of central frequency unit (CPU) cycles, denoted by $c_k > 0$.
Furthermore, we assume that the computation results or the task-output bits have much smaller size than the task-input bits, and hence the computation downloading time from GBSs to the UAV is practically negligible and thus omitted.

Consider a 3D Cartesian coordinate system, in which each GBS $k\in\mathcal K$ has zero altitude and fixed horizontal location ${\bm \nu}_k=(x_k,y_k)$.
We assume that the UAV flies at a fixed altitude $H\geq0$ in meter (m), and use $\mv u_I=(x_I,y_I)$ and $\mv u_F=(x_F,y_F)$ to denote the UAV's initial and final locations projected on the horizontal plane, respectively.
Furthermore, we denote the mission completion time as $T$ in second (s), which is a variable to be optimized later.
Let $\hat{\mv u}(t)=(\hat{x}(t),\hat{y}(t))$ denote the UAV's horizontal location at time instant $t\in[0,T]$. Then we have $\hat{\mv u}(0)={\bm u}_I$ and $\hat{\mv u}(T)={\bm u}_F$ for the given initial and final locations, respectively. At time instant $t$, the distance between the UAV and GBS $k$ is given by
\begin{align}\label{UAV_GBS_dis}
d_k(\hat{\mv u}(t))=\sqrt{H^2+\|\hat{\mv u}(t)-{\bm \nu}_k\|^2},
\end{align}
where $\|\cdot\|$ denotes the Euclidean norm of a vector. Let $V_{\max}>0$ denote the UAV's maximum speed in m/s. Then we have $\sqrt{\dot{\hat{x}}^2(t)+\dot{\hat{y}}^2(t)}\leq V_{\max}$, $\forall t\in[0,T]$, in which $\dot{\hat{x}}(t)$ and $\dot{\hat{y}}(t)$ denote the first-derivatives of $\hat{x}(t)$ and $\hat{y}(t)$, respectively.

Normally, the air-to-ground channels from the UAV to GBSs are dominated by the LoS links, and hence we consider the free-space path-loss model similarly as in \cite{shuowenzhang17,Huangyw17}. At time instant $t$, the channel power gain from the UAV to GBS $k$ is denoted as
\begin{align}\label{UAV_GBS_h(t)}
h_k(\hat{\mv u}(t))=\frac{\beta_0}{d_k^2(\hat{\mv u}(t))}=\frac{\beta_0}{H^2+\|\hat{\mv u}(t)-{\bm \nu}_k\|^2},
\end{align}
where $\beta_0$ denotes the channel power gain at a reference distance of $1$~m.

For ease of exposition, we discretize the mission duration $T$ into $N$ time slots each with a given duration $\delta_t$, i.e. $T=N\delta_t$, where  $\delta_t$ is chosen to be sufficiently small such that the UAV's location can be assumed to be approximately unchanged during each slot with $\delta_tV_{\max}\ll H$, and $N$ is thus a variable to be optimized. In this case, we denote the UAV's horizontal location at time slot $n$ as ${\bm u}[n]\triangleq \hat{\mv u}(n\delta_t),n\in {\cal N}\triangleq\{1,...,N\}$, with ${\bm u}[0]\triangleq \hat{\mv u}(0)={\bm u}_I$ and ${\bm u}[N] \triangleq \hat{\mv u}(T) = {\bm u}_F$. Accordingly, the channel power gain from the UAV to GBS $k$ is $h_k({\bm u}[n])$ at slot $n$. Furthermore, let $S_{\max}=\delta_tV_{\max}$ denote the maximum UAV displacement during each time slot. Thus, the maximum UAV speed and initial/final location constraints are respectively re-expressed as
\begin{align}
\|{\bm u}[n]-{\bm u}[n-1]\|^2\le S_{\max}^2,\forall n \in\mathcal N,\label{eqn:speed}\\
{\bm u}[0]={\bm u}_I,~{\bm u}[N] = {\bm u}_F.\label{eqn:inifin}
\end{align}

We consider the time-division-multiple-access (TDMA) protocol to implement the UAV's computation offloading, by dividing each time slot $n \in\mathcal N$ into $K$ sub-slots each with duration $\tau_k[n]\geq 0$, where
\begin{align}
 \sum \limits_{k\in{\cal K}} \tau_k[n]=\delta_t, ~\forall n \in\mathcal N. \label{eqn:tdma}
\end{align}
In each sub-slot $k \in \mathcal K$, the UAV offloads the respective task-input bits to GBS $k$.
Suppose that the UAV adopts a constant transmit power $P > 0$ for offloading.
Then the achievable offloading rate from the UAV to GBS $k$ in bits-per-second (bps) at slot $n$ is expressed as
\begin{align}\label{eq:R[n]}
R_k({\bm u}[n])&=B\log_2\left(1+\frac{Ph_k(\mv u[n])}{\sigma^2}\right), \notag\\
&=B\log_2\left(1+\frac{\rho }{H^2+\|{\bm u}[n]-{\bm \nu}_k\|^2}\right),
\end{align}
where $\sigma^2$ and $B$ represent the noise power at the receiver of each GBS and the bandwidth, respectively, and $\rho=\frac{P\beta_0}{\sigma^2}$ denotes the reference signal-to-noise ratio (SNR).
In order for the UAV to offload all the $L$ task-input bits to the $K$ GBSs, we need to have
\begin{align}\label{L}
\sum \limits_{k\in{\cal K}} \sum \limits_{n\in{\cal N}} \tau_k[n]R_k({\bm u}[n]) \geq L.
\end{align}

Next, we consider the remote task execution at each GBS $k$. Denoting $f_k$ as the maximum CPU frequency at GBS $k\in\mathcal K$ in Hz, then we obtain the per-slot computation capacity of GBS $k$ as $f_k\delta_t$, which represents the maximum number of task-input bits that can be executed by GBS $k$ over one slot. Note that as each task-input bit can be executed independently, each GBS can immediately start the execution as soon as the task-input bits are received. In other words, the offloaded task-input bits at each slot $n-1$ are immediately executable at slot $n$. Also note that at each GBS $k$, all the offloaded task-input bits must be successfully executed before the mission completion time $T$ (or $N$). Therefore, we have the following computation capacity constraints over time: for each GBS $k\in \mathcal K$, the accumulative number of offloaded task-input bits over the last $(N-n+1)$ slots must be no larger than the GBS's accumulative computation capacity over the last $(N-n)$ slots, $\forall n\in\mathcal N$, i.e.,
\begin{align}\label{eq:ln:max}
\sum_{j=n}^N c_k\tau_k[j]R_k({\bm u}[j]) \leq (N-n)f_{k}\delta_t, \forall n\in\mathcal N.
\end{align}
The computation capacity constraints in \eqref{eq:ln:max} can be understood intuitively as follows. First, for $n=N$, we have $c_k\tau_k[N]R_k({\bm u}[N]) = 0$, which indicates that the UAV cannot offload any task in slot $N$, as there is no time for each GBS to execute. Next, for $n=N-1$, we have $c_k\tau_k[N-1]R_k({\bm u}[N-1]) + c_k\tau_k[N]R_k({\bm u}[N]) \le f_k \delta_t$. By combining this with $c_k\tau_k[N]R_k({\bm u}[N]) =0$, we further have $c_k\tau_k[N-1]R_k({\bm u}[N-1]) \le f_k \delta_t$, which implies that the offloaded task-input bits in slot $N-1$ cannot exceed the computation capacity in slot $N$. Furthermore, by recursively considering time slots $N-2,N-3,\ldots$, until the first slot, the constraints in \eqref{eq:ln:max} follow similarly.


Our objective is to minimize the UAV's mission completion time $N$ (or equivalently $T$) by optimizing the UAV trajectory $\{ {\bm u}[n]\}$ and the time allocation for computation offloading $\{\tau_k[n]\}$, subject to the maximum UAV speed constraint in \eqref{eqn:speed}, the initial/final UAV location constraints in \eqref{eqn:inifin}, the TDMA constraints in \eqref{eqn:tdma}, as well as the task execution constraints in \eqref{L} and \eqref{eq:ln:max}.
Therefore, the joint UAV trajectory and computation offloading optimization problem is formulated as
\begin{align}
\mathbf{(P1):} \min_{\{ {\bm u}[n],\tau_k[n]\},N\in \mathbb{Z}^+} &~~N\notag\\
{\rm s.t.}~~~~&  \tau_k[n]\geq0,~\forall k\in \mathcal K, n\in \mathcal N \label{eqn:P:ak}\\
~&  \eqref{eqn:speed},~\eqref{eqn:inifin},~\eqref{eqn:tdma},~ \eqref{L},~{\text{and}}~ \eqref{eq:ln:max},\notag
\end{align}
where $\mathbb{Z}^+$ denotes the set of all strictly positive integers.
Notice that (P1) is a non-convex optimization problem, as the optimization variable $N$ is an integer, and constraints \eqref{L} and \eqref{eq:ln:max} are non-convex. Furthermore, as $N$ is {\it a-priori} unknown, (P1) consists of an uncertain number of constraints in \eqref{eqn:speed}, \eqref{eqn:tdma}, and \eqref{eq:ln:max}. Due to the above facts, (P1) is difficult to be solved optimally.

\vspace{-0.2cm}
\section{Proposed Solution to (P1)}\vspace{-0.2cm}
In this section, we propose an efficient algorithm to solve (P1) sub-optimally.

First, we show that (P1) can be equivalently solved by first optimizing over $\{ {\bm u}[n]\}$ and $\{\tau_k[n]\}$ under any given $N$, and then using a bisection search to find the optimal $N$. In particular, under any given $N$, (P1) becomes the following feasibility checking problem:
\begin{align*}
\mathbf{(P2):}~{\rm find} ~&\{{\bm u}[n]\}~{\text{and}}~ \{\tau_k[n]\} \\
{\rm s.t.}
~&  \eqref{eqn:speed},~\eqref{eqn:inifin},~\eqref{eqn:tdma},~ \eqref{L},~ \eqref{eq:ln:max},~{\text{and}}~\eqref{eqn:P:ak}.\notag
\end{align*}
Suppose that the optimal solution of $N$ to (P1) is $N^\star$. Then, consider (P2) under any given $N$. If (P2) is feasible under $N$, then it follows that $N^\star \le N$; otherwise, we have $N^\star > N$. Therefore, we can solve (P1) by checking the feasibility of (P2) under any given $N$ and using a bisection search over $N$. As a result, we only need to consider (P2) under given $N$.


Next, we show that solving (P2) is equivalent to solving the following problem (P3) to maximize the number of computation task-input bits under given $N$.
\begin{align}
\mathbf{(P3):} \max_{\{{\bm u}[n]\},\{\tau_k[n]\},\tilde{L}\geq0} &~~\tilde{L}\notag\\
{\rm s.t.}~&\sum \limits_{k\in{\cal K}} \sum \limits_{n\in{\cal N}} \tau_k[n]R_k({\bm u}[n]) \geq \tilde{L}, \label{eqn:(P3):L}\\
~&  \eqref{eqn:speed},~\eqref{eqn:inifin},~\eqref{eqn:tdma},~ \eqref{eq:ln:max},~{\text{and}}~\eqref{eqn:P:ak}.\notag
\end{align}
Suppose that the optimal solution of $\tilde{L}$ to (P3) is $\tilde{L}^*$. Then it is evident that if $\tilde{L}^* \geq L$, then (P2) is feasible; otherwise, (P2) is infeasible.

Now, it only remains to solve (P3). Note that (P3) is still non-convex, due to the non-convex constraints in \eqref{eq:ln:max} and \eqref{eqn:(P3):L}. In the following, we propose an efficient algorithm to obtain a suboptimal solution to (P3) by optimizing the time allocation $\{\tau_k[n]\}$ and the UAV trajectory $\{{\bm u}[n]\}$ in an alternating manner.


\subsubsection{Time Allocation for (P3) Under Given UAV Trajectory}
Under given $\{ {\bm u}[n]\}$, (P3) is reduced to
\begin{subequations}
\begin{align}
\mathbf{(P3.1):} \max_{\{\tau_k[n]\},\tilde{L}\geq0} \tilde{L}
~~{\rm s.t.}~\eqref{eqn:tdma},~\eqref{eq:ln:max},~ \eqref{eqn:P:ak},~{\text{and}}~\eqref{eqn:(P3):L}.\notag
\end{align}
\end{subequations}
It is easy to show that (P3.1) is a linear program (LP), which can be solved by standard convex optimization techniques such as the interior point method \cite{Boyd2004}. We adopt the well-established optimization toolbox CVX \cite{cvx} to solve (P3.1) optimally and efficiently. 
\subsubsection{UAV Trajectory Optimization for (P3) Under Given Time Allocation}
Under given $\{\tau_k[n]\}$, (P3) is reduced to
\begin{align}
\mathbf{(P3.2):}\max_{\{ {\bm u}[n]\},\tilde{L}\geq0} \tilde{L}
~~{\rm s.t.}~\eqref{eqn:speed},~\eqref{eqn:inifin},~ \eqref{eq:ln:max},~{\text{and}}~\eqref{eqn:(P3):L}.\notag
\end{align}
Notice that (P3.2) is still non-convex, as constraints \eqref{eq:ln:max} and \eqref{eqn:(P3):L} are non-convex. To tackle this problem, we propose an iterative algorithm to obtain an efficient solution to (P3.2) by using the SCA technique. The idea is that under any given local point at each iteration, we approximate non-convex constraints \eqref{eq:ln:max} and \eqref{eqn:(P3):L} by their corresponding convex ones. By solving a series of approximate convex problems iteratively, we can attain an efficient suboptimal solution to (P3.2).

Suppose that $\{{\bm u}^{(i)}[n]\}$ denotes the local point at the $i$-th iteration, $i \ge 0$. Then, we approximate constraints \eqref{eq:ln:max} and \eqref{eqn:(P3):L} in the following, respectively. First, consider constraint \eqref{eq:ln:max}. Notice that by checking the first-order Taylor expansion of the convex term $H^2+\|{\bm u}[n]-{\bm \nu}_k\|^2$ with respect to ${\bm u}[n]$ at the local point ${\bm u}^{(i)}[n]$, we have
\begin{align}\label{p3_up_d}
H^2+\|{\bm u}[n]-{\bm \nu}_k\|^2 \geq q_k^{(i)}[n]+ 2({\bm \omega}^{(i)}[n])^T{\bm u}[n],
\end{align}
with ${\bm \omega}^{(i)}[n]={\bm u}^{(i)}[n]-{\bm \nu}_k$ and $q_k^{(i)}[n]=H^2+\|{\bm u}^{(i)}[n]-{\bm \nu}_k\|^2-2({\bm \omega}^{(i)}[n])^T{\bm u}^{(i)}[n]$, where $(\cdot)^T$ indicates the transpose. Based on \eqref{p3_up_d}, we obtain an upper bound of $R_k({\bm u}[n])$ as
\begin{align*}
R_k({\bm u}[n])&\leq B\log_2\left(1+\frac{\rho}{q_k^{(i)}[n]+ 2({\bm \omega}^{(i)}[n])^T{\bm u}[n]}\right)\\
&\triangleq R_{k,{\rm up}}^{(i)}({\bm u}[n]),
\end{align*}
where $R_{k,{\rm up}}^{(i)}({\bm u}[n])$ is convex with respect to ${\bm u}[n]$.
Replacing $R_k({\bm u}[n])$ in \eqref{eq:ln:max} as $R_{k,{\rm up}}^{(i)}({\bm u}[n])$, we have the approximated convex constraints as
\begin{align}\label{p3_f/c}
\sum_{j=n}^{N} c_k\tau_k[n]R_{k,{\rm up}}^{(i)}({\bm u}[n])\leq (N-n) f_{k}\delta_t, ~\forall n\in \mathcal N.
\end{align}

Next, consider constraint \eqref{eqn:(P3):L}. Notice that $R_k({\bm u}[n])$ is a convex function with respect to the term $ \|{\bm u}[n]-{\bm \nu}_k\|^2$. Then by taking the first-order Taylor expression of $R_k({\bm u}[n])$ with respect to $\|{\bm u}[n]-{\bm \nu}_k\|^2$, we can obtain a lower bound of $R_k({\bm u}[n])$
at local point ${\bm u}^{(i)}[n]$ as follows.
\begin{align}
&R_k({\bm u}[n])\geq R_{k,\text{low}}^{(i)}({\bm u}[n])\notag\\
&\triangleq R_k({\bm u}^{(i)}[n])-b_k^{(i)}[n](\|{\bm u}[n]-{\bm \nu}_k\|^2-\|{\bm u}^{(i)}[n]-{\bm \nu}_k\|^2),\label{R_low}
\end{align}
where $b_k^{(i)}[n]=B\rho/(\ln2d_k^2({\bm u}^{(i)}[n])(\rho+d_k^2({\bm u}^{(i)}[n])))$. Here, $R_{k,\text{low}}^{(i)}(\bm u[n])$ is a concave function with respect to ${\bm u}[n]$. By replacing $R_k({\bm u}[n])$ in constraint \eqref{eqn:(P3):L} as $R_{k,\text{low}}^{(i)}({\bm u}[n])$, we have the approximated convex constraints as
\begin{align}\label{p3_tilde_L}
\sum \limits_{k\in{\cal K}} \sum \limits_{n\in{\cal N}}  \tau_k[n]R_{k,\text{low}}^{(i)}(u[n]) \geq \tilde{L}.
\end{align}

Finally, with \eqref{p3_f/c} and \eqref{p3_tilde_L} at hand, (P3.2) is approximated as the following convex optimization problem (P3.3) at local point $\{{\bm u}^{(i)}[n]\}$, which can be solved optimally via convex optimization techniques such as CVX.
\begin{align}
\mathbf{(P3.3):}\max_{\{ {\bm u}[n]\},\tilde{L}\geq0} \tilde{L}
~~{\rm s.t.}~&\eqref{eqn:speed},~\eqref{eqn:inifin},~ \eqref{p3_f/c},~{\text{and}}~\eqref{p3_tilde_L}.\notag
\end{align}
Let $\{{\bm u}^{(i)*}[n]\}$ denote the optimal UAV trajectory solution to (P3.3) at local point $\{{\bm u}^{(i)}[n]\}$. Then, we can obtain an efficient iterative algorithm to solve (P3.2) as follows. In each iteration $i \ge 1$, the UAV trajectory is updated as $\{{\bm u}^{(i)*}[n]\}$ by solving (P3.3) at local point $\{{\bm u}^{(i)}[n]\}$, i.e. ${\bm u}^{(i+1)}[n]={\bm u}^{(i)*}[n],\forall n\in\mathcal N$, where $\{{\bm u}^{(0)}[n]\}$ denotes the initial UAV trajectory. In summary, the proposed algorithm is presented in Table \uppercase\expandafter{\romannumeral1} as Algorithm 1.
\begin{table}[htp]\vspace{-0.3cm}
\begin{center}\label{algorithmfor(P3.2)}
\caption{Algorithm 1 for Solving Problem (P3.2)}  \vspace{0.1cm}
\hrule \vspace{0cm} 
    \begin{itemize}
    \item[1]  Initialization: Given the UAV trajectory $\{{\bm u}^{(0)}[n]\}$; let $i=0$.
    \item[2]  {\bf Repeat:}
                \begin{itemize}
                \item[i]  Solve problem (P3.3) under given $\{{\bm u}^{(i)}[n]\}$ to obtain the optimal solution as $\{{\bm u}^{(i)*}[n]\}$.
                 \item[ii]  Update ${\bm u}^{(i+1)}[n]={\bm u}^{(i)*}[n],\forall n\in\mathcal N$.
                \item[iii]   Update $i=i+1$.
                \end{itemize}
     \item[3] {\bf Until} the optimal value converges within a given threshold or a maximum number of iterations is reached.
    \end{itemize}
\hrule \vspace{0cm}
\end{center}\vspace{-0.5cm}
\end{table}

Notice that after each iteration in Algorithm 1, the objective value of (P3.2) is monotonically non-decreasing. As the optimal value of (P3.2) is upper-bounded, Algorithm 1 should converge to (at least) a locally  optimal solution to (P3.2).

\subsubsection{Complete Algorithm to Solve (P3)}
With (P3.1) and (P3.2) solved, we are ready to solve (P3) by updating the UAV trajectory $\{ {\bm u}[n]\}$ and time allocation $\{\tau_k[n]\}$ in an alternating manner. In each iteration, we first solve (P3.1) under given $\{ {\bm u}[n]\}$ to update $\{\tau_k[n]\}$, and then solve (P3.2) under $\{\tau_k[n]\}$ to update $\{ {\bm u}[n]\}$. For each iteration, the optimal value of (P3) is monotonically nondecreasing. As the optimal value of (P3) is upper-bounded, the alternating-optimization-based algorithm will converge to at least a locally optimal solution to (P3).

Finally, with (P3) solved, the feasibility of (P2) is accordingly checked. By combing this together with the bisection search over $N$, problem (P1) can be efficiently solved. Here, it is worth noting that the obtained solution to (P1) is generally suboptimal, which is due to the fact that under given $N$, we only obtain a locally optimal solution to (P3). Nevertheless, as shown in numerical results next, such suboptimal solution to (P1) performs quite well in practice.


\vspace{-0cm}
\section{Numerical Results}
\begin{figure}
\centering
 \setlength{\abovecaptionskip}{-2.5mm}
\setlength{\belowcaptionskip}{-2.5mm}
    \includegraphics[width=3in]{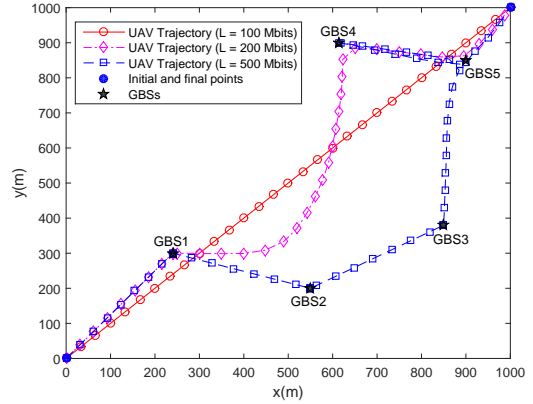}
\caption{Optimized UAV trajectory projected on the horizontal plane under different values of $L$. } \label{fig:3}
\vspace{-0.8cm}
\end{figure}

\vspace{-0cm}
In this section, we present numerical results to validate the proposed joint trajectory and computation offloading design. Suppose that there are $K=5$ GBSs that are distributed within a geographic area of size $1\times 1$~km$^2$, as shown in Fig.~\ref{fig:3}. We set the bandwidth as $B=1$~MHz and the flying altitude of the UAV as $H=50$~m. The channel power gain at the reference distance of $1$~m is $\beta_0=-30$~dB and the noise power at each GBS receiver is $\sigma^2=-60$~dBm. The maximum UAV speed is $V_{\max}=50$~m/s while the transmit power is $P=30$ dBm. At each GBS $k\in\mathcal K$, we set the maximum CPU frequency as $f_k=2.5$~GHz and the required number of CPU cycles per task-input bit as $c_k=10^3$.
Furthermore, the initial UAV trajectory for Algorithm 1 is heuristically
designed as follows.
\begin{itemize}
\item {\it Straight flight:} the UAV flies straight from the initial to the final location at a fixed speed $V = \|{\bm u}_F - {\bm u}_I\|/T$. To minimize the mission completion time under this trajectory, we first check the computation feasibility under any given $T$ (or $N$) by optimizing the time allocation, and then use a bisection search over $T$ (or $N$).
\end{itemize}

Fig.~\ref{fig:3} shows the optimized UAV trajectory projected on the horizontal plane under different values of $L$, in which the trajectory is sampled every $\delta_t=1$ s. It is observed that when $L=100$~Mbits, the UAV flies straight from the initial location to the final location at the maximum speed, and the mission completion time is $T=\|{\bm u}_F-{\bm u}_I\|/V_{\max}$, which is constrained by the flying distance between the initial and final locations. When $L = 200$ Mbits, the UAV trajectory is observed to deviate from the straight line by flying closer towards GBSs 1, 4, and 5, in order to exploit better wireless channels for computation offloading towards them. When $L$ further increases to $500$ Mbits, the UAV is observed to reach and hover above all the five GBSs and even fly back and forth between GBSs 4 and 5. In this case, the mission completion time is mainly constrained by the computation task execution, and thus the UAV trajectory is designed for most efficient computation offloading.

Next, we validate the performance of our proposed design as compared to two benchmark schemes, namely the above straight flight trajectory and the following heuristic design.
\begin{itemize}
\item {\it Successive hover-and-fly:} the UAV flies to successfully reach at the top of the $K$ GBSs at the maximum speed $V_{\max}$, and hovers above each of them for efficient computation offloading. The visiting order is determined by solving the Traveling Salesman Problem (TSP) \cite{tsp} to minimize the flying distance. Under such a UAV trajectory design, the mission completion time minimization problem can be solved similarly as in the straight flight scheme, while the only difference is that during checking the computation feasibility under any given $T$ (or $N$), we need to optimize the hovering durations above these GBSs jointly with the time allocation while flying. 
\end{itemize}

\begin{figure}
\centering
 \setlength{\abovecaptionskip}{-2.5mm}
\setlength{\belowcaptionskip}{-2.5mm}
    \includegraphics[width=3in]{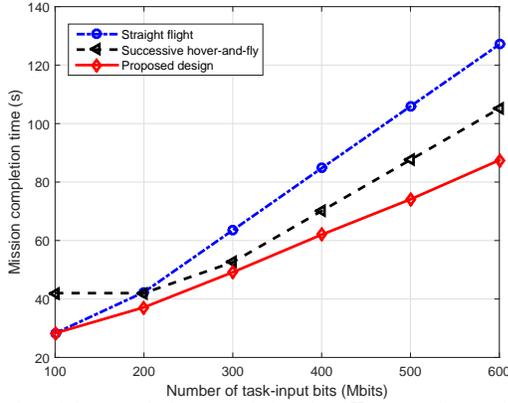}
\caption{The minimum mission completion time $T$ versus the number of task-input bits $L$. } \label{fig:4}
\vspace{-0.8cm}
\end{figure}
Fig.~\ref{fig:4} shows the mission completion time $T = N\delta_t$ versus the number of task-input bits $L$.
It is observed that as $L$ becomes larger, the mission completion time increases for all the three schemes, while the proposed design performs best among the three schemes over all $L$ values.
When $L$ is small (e.g., $L=100$ Mbits), the straight-flight scheme is observed to achieve the same mission completion time as the proposed design, and outperforms the successive-hover-and-fly scheme. This is due to the fact that in this regime, the mission completion time is constrained by the flying distance, and the successive-hover-and-fly scheme leads to longer flying distance as the UAV needs to visit all GBSs. When $L$ is larger than $200$ Mbits, it is observed that the straight-flight scheme performs worse than the successive-hover-and-fly scheme and the proposed design. This is due to the fact that in this regime, the mission completion time is constrained by the computation task execution, and the latter two schemes can more efficiently explore the UAV trajectory design for computation offloading. In addition, the successive-hover-and-fly scheme is observed to perform close to the proposed design when $L=300$ Mbits, but the performance gap increases when $L$ further increases. This is due to the fact that when $L$ becomes larger, in the proposed design the UAV can fly back and forth among different GBSs (see Fig.~2 for $L = 500$ Mbits) in order to explore multiple GBSs' distributed computation resources more efficiently by time sharing.

\vspace{-0cm}
\section{Conclusion}\vspace{-0cm}
This paper investigates a new MEC application scenario where a cellular-connected UAV offloads its computation tasks to multiple GBSs along its trajectory. The UAV trajectory is jointly designed with the computation offloading scheduling, to minimize the mission completion time, subject to the UAV's maximum speed and initial/final location constraints, as well as the GBSs' individual computation capacity constraints. By exploiting alternating optimization and SCA techniques, an efficient algorithm is proposed to solve the formulated problem sub-optimally. Numerical results show a significant performance gain of our proposed design over the benchmark schemes.

\end{document}